\documentclass[preprint]{ptephy_v1}

\usepackage{boites}
\usepackage{amsmath}
\usepackage{amsthm}
\usepackage{mathtools}
\usepackage{mathrsfs}
\usepackage{color}
\usepackage{braket}
\usepackage{url}
\usepackage{bm}
\usepackage{hyperref}

\preprintnumber{J-PARC-TH-0240, KUNS-2860}

\usepackage[normalem]{ulem}  

\renewcommand\sout{\bgroup\color[rgb]{1,0.75,0.8} \ULdepth=-.5ex \ULset}

\begin{document}

\title{Beta-decay formulas revisited (I):
  Gamow--Teller and spin-dipole contributions to
allowed and first-forbidden transitions}

\author{Wataru Horiuchi}
\affil[1]{Department of Physics, Hokkaido University, Sapporo 060-0810, Japan}

\author[2,3]{Toru Sato}
\affil[2]{Research Center for Nuclear Physics, Osaka University, Ibaraki, Osaka 567-0047, Japan} 
\affil[3]{J-PARC Branch, KEK Theory Center,
     Institute of Particle and Nuclear Studies (KEK) 
     and Theory Group, Particle and Nuclear Physics Division, J-PARC Center,
     Tokai, Ibaraki, 319-1106, Japan}
     
\author[4]{Yuichi Uesaka}
\affil[4]{Faculty of Science and Engineering, Kyushu Sangyo University, Fukuoka 813-8503, Japan}

\author[5]{Kenichi Yoshida}
\affil[5]{Department of Physics, Kyoto University, Kyoto 606-8502, Japan}

\begin{abstract}
We propose formulas of the nuclear beta-decay rate that are useful in a practical calculation. 
The decay rate is determined by the product of the lepton and hadron current densities. 
A widely used formula relies upon the fact that the low-energy lepton wave functions in a nucleus 
can be well approximated  
by a constant and linear to the radius for the $s$-wave and $p$-wave wave functions, respectively. 
We find, however, the deviation from such a simple approximation is
evident for heavy nuclei with large $Z$ 
by numerically solving the Dirac equation. 
In our proposed formulas, 
the neutrino wave function is treated exactly as a plane wave, while the electron wave function is 
obtained by iteratively solving the integral equation, thus we can control the uncertainty of the approximate 
wave function. 
The leading-order approximation gives a formula  equivalent to the
conventional one and overestimates the decay rate. 
We demonstrate that the next-to-leading-order formula reproduces well the exact result for a 
schematic transition density as well as a microscopic one obtained by a nuclear energy-density functional method.

\end{abstract}

\maketitle
\allowdisplaybreaks[1]

\section{Introduction}

The physics of exotic nuclei away from the stability line has been a major subject in nuclear physics. 
The lifetime of neutron-rich nuclei is governed by beta decay.
Since the beta decay determines the time scale of the rapid-neutron-capture process ($r$-process) 
and the production of heavy elements 
together with the beta-delayed neutron(s) emission, 
the beta-decay rates of exotic nuclei are an important microscopic input for the simulation of nucleosynthesis~\cite{lan03}. 
The multi-messenger observations from a binary neutron star merger~\cite{abb17a, abb17b} imply that 
heavy neutron-rich nuclei that are even close to the drip line are involved in the $r$-process.
Thus, the Coulomb effect on the beta particle (emitted electron) should be carefully examined under 
the extreme environment where
the $Q$ value for the beta decay, $Q_\beta$, is high and the nuclear charge $Z$ is large.

A careful analysis of the Coulomb effect is also useful
for a precision test
of the standard model to find a signal of
  new physics.
  For example, the effect on spectra of the beta particle
  and angular correlation as well as beta-decay rates
  has been studied to test the unitarity of the
  Cabibbo--Kobayashi--Maskawa (CKM) matrix, 
the scalar and tensor interactions, and the effect of neutrino mass in the allowed and
first-forbidden transitions
~\cite{Ando:2009jk,Glick-Magid:2016rsv,Gonzalez-Alonso:2018omy,Cirgiliano:2019nyn}.

  The formulation of nuclear beta decay
  within the distorted-wave impulse approximation of the electron
  Coulomb interaction has been matured
  ~\cite{stech1964nuclear,schulke1964nuclear,Behrens:1971rbq,sch66,Schopper:1969jkp,Morita:1963zz,morita1973beta,Koshigiri:1979zd}.
  The crucial part is how to handle  the electron Coulomb wave function
  with a potential
  of the finite-size nuclear-charge distribution.
  Using the Maclaurin expansion of the nuclear radius $r$, the exact
  electron wave function
  was included in Ref. \cite{buhring1963beta}.
  An iterative solution of the integral equation was found to have
  a better convergence
  by Behrens and B\"{u}hling~\cite{rose1951note,Behrens:1971rbq}.
The formula is arranged
in the order of ${\cal O}(r^a V_C^b E_e^c m_e^d)$, where
$V_C$, $E_e$ and $m_e$ represent the Coulomb potential, the energy and the mass of an electron, respectively.
It has been widely used in the calculations 
such as in Refs.~\cite{gov71,war88,war91} and in the recent
application to the $r$-process nuclei
~\cite{eng99,mol03,bor00,bor03,bor06,cue07,suz12,niu12,zhi13,mus16,mar16,ney20}.
In most of those works, however,
the leading-order approximation of the formula
in Refs.~\cite{Behrens:1971rbq,Schopper:1969jkp} is adopted.
Instead of expanding the lepton wave functions,
one can incorporate the numerical solution of the charged lepton wave functions
thanks to the advance of the computational ability.
The muon capture~\cite{morita1960theory}
and the beta decay~\cite{Koshigiri:1979zd} are
formulated suitable for this purpose.
In this formulation, the nuclear matrix element is defined in a transparent way
and
  appears similarly in Refs.~\cite{Nakamura:2000vp,walecka2012semileptonic}
for the semi-leptonic nuclear processes and
electron scattering~\cite{DeForest:1966ycn}. It is thus
straightforward to apply it to the charged-current neutrino reaction
and lepton capture reaction. Developing an analytic formula of beta decay
based on Ref.~\cite{Koshigiri:1979zd} would also contribute
to a precise understanding of the neutrino-nucleus reactions
to extract neutrino properties from neutrino experiments
as discussed in Ref.~\cite{nak17,Alvarez-Ruso:2017oui}.

The high-energy forbidden transitions occur 
under the exotic environment with high $Q_{\beta}$ value~\cite{yos17}.
Therefore, in this work we revisit the formulation of beta decay
for not only the
allowed but the first-forbidden transitions induced by the Gamow--Teller and spin-dipole
type operators.
We provide a simple way to improve the widely used formula in the
nuclear beta-decay study to  apply for nuclei with 
large $Z$ and away from the stability line.
We start from the formulation of 
Koshigiri {\it et al}.~\cite{Koshigiri:1979zd} and use iterative solutions of the integral
equation~\cite{rose1951note,Behrens:1971rbq}.
In the previous formalism, one often expands the electron
and neutrino wave
functions in the long wave-length approximation and collect 
terms in a systematic way.
Here we avoid this expansion of the neutrino wave function.
We use an analytic form of the
LO and NLO electron wave functions combined with the numerical table of the 
electron wave function at the origin.
This makes the formula of the beta-decay rate simple and easy to use.

This paper is organized in the following way. 
We start from the formulation of 
the beta decay with the partial wave expansion for the lepton wave functions
in Sect.~\ref{formalism}.
We provide an explicit expression of the first (LO) and the second (NLO) iteration
of the integral equation for an electron wave function in Sect.~\ref{WF}. 
Formulas of the beta-decay rate are given and compared with the widely used one
in Sect.~\ref{formula}.
The formulas of LO and NLO are examined in Sect.~\ref{model},
using a schematic transition density that is given by a sum of two Gaussians. 
We then in Sect.~\ref{EDF} apply the formulas to the neutron-rich Ni and Sn isotopes where 
the transition densities are microscopically obtained by a nuclear 
energy-density functional (EDF) method. 
Summary and perspectives are given finally in Sect.~\ref{summary}.

\section{Formalism}\label{formalism}

\subsection{Effective Hamiltonian}

An effective Hamiltonian for a low-energy charged-current reaction
is given as
\begin{eqnarray}
  H_{\mathrm{eff}} = \frac{G_F V_{ud}}{\sqrt{2}}
  \int d\bm{x}[
    \bar{\ell}(\bm{x})\gamma^\mu(1-\gamma_5)\nu_\ell(\bm{x}) J_\mu(\bm{x})
    + \bar{\nu}_\ell(\bm{x})\gamma^\mu(1-\gamma_5)\ell(\bm{x})J_\mu^\dagger(\bm{x})],
    \label{eq:effective_hamiltonian}
\end{eqnarray}
where $\ell(\bm{x})$ represents either the electron, muon, or tau field and $\bar{\psi}=\psi^\dagger \gamma^0$.
The hadron current $J^\mu(\bm{x})$ is given by
the vector  and axial vector currents
\begin{eqnarray}
  J^\mu(\bm{x}) = V^\mu(\bm{x}) - A^\mu(\bm{x}),
  \label{eq:hadron_current}
\end{eqnarray}
where $G_{\mathrm{F}}=1.166\times 10^{-5}\mathrm{GeV}^{-2}$
  is the Fermi coupling constant and $V_{ud}=0.9737$ is
  the CKM matrix~\cite{Zyla:2020zbs}. Here we take natural units
  $\hbar=c=1$.

The effective Hamiltonian describes semi-leptonic nuclear weak processes
such as lepton capture, neutrino reaction, 
and $\beta^{\pm}$ decay. 
For $\beta^-$ decay,
$ i \rightarrow e^-(p_e) +\bar{\nu}_e(p_\nu) + f$,
where $i$ and $f$ respectively denote
the initial and final nuclear states,
and $p_\ell$ is the  lepton momentum,  
the transition matrix element is given as
\begin{align}
   \left<e^-(p_e)\bar{\nu}_e(p_\nu) f\right|H_{\mathrm{eff}}\left|i\right>
 = \frac{G_F V_{ud}}{\sqrt{2}}  \int d\bm{x}
  \bar{\psi}_{e^-,p_e,s_e}^{(-)}(\bm{x})\gamma^\mu (1-\gamma_5)v_{s_\nu}(\bm{p}_\nu)
  e^{-i\bm{p}_\nu\cdot\bm{x}}\left<f\right|J_\mu(\bm{x})\left|i\right>,
  \label{eq:transition_amplitude_beta_decay}
\end{align}
and for $\beta^+$ decay
\begin{align}  
\left<e^+(p_e)\nu_e(p_\nu)f\right|H_{\mathrm{eff}}\left|i\right>
 = \frac{G_F V_{ud}}{\sqrt{2}}  \int d\bm{x}
\bar{u}_{s_\nu}(\bm{p}_\nu)  e^{-i\bm{p}_\nu\cdot\bm{x}}\gamma^\mu (1-\gamma_5)
   \psi_{e^+,p_e,s_e}^{(+)}(\bm{x})
\left<f\right| J_\mu^\dagger(\bm{x})\left|i\right>,
  \end{align}
where $u$ and $v$ are
the Dirac spinors of the neutrino and antineutrino, respectively.
 The electron scattering wave functions $\psi$ 
with the superscript $(-)$ and $(+)$
satisfy the incoming and outgoing boundary conditions, respectively.

\subsection{Multipole Expansion of the Effective Hamiltonian}

The standard formulation of the
beta decay adopts the partial wave expansion of
both neutrino and electron wave functions. 
We use the following  electron
(charged lepton in general) scattering wave function
\begin{align}
  \psi_{e,p_e,s_e}^{(\mp)}(\bm{x})
   =  \sum_{\kappa_e,m_e,\mu_e}(4\pi) i^{l_{\kappa_e}}(l_{\kappa_e},m_e,1/2,s_e|j_{\kappa_e},\mu_e)Y^*_{l_{\kappa_e},m_e}(\hat{p}_e)
  e^{\mp i\Delta_{\kappa_e}}
  \left ( \begin{array}{c}
    G_{\kappa_e}(r)\chi_{\kappa_e}^{\mu_e} \\
  i F_{\kappa_e}(r) \chi_{-\kappa_e}^{\mu_e}
  \end{array}
  \right ).
\end{align}
Here, $(j_1,m_1,j_2,m_2|J,M)$ is the Clebsch--Gordan coefficient
  \cite{de1974nuclear,rose1957ang,condon1935theory}.
For positron, $G_{\kappa_e}, F_{\kappa_e}$  are calculated 
by replacing $Z$ of the Coulomb interaction by $- Z$.
$\Delta_{\kappa_e}$ is the Coulomb phase.
The normalization of the scattering wave function 
in the plane wave expansion
$\psi_e(\bm{x}) \rightarrow u(p_e)\exp(i\bm{p}_e\cdot\bm{x})$ 
is given as
\begin{align}
  u_{s_e}(p_e) = \sqrt{\frac{E_e+m_e}{2E_e}}\left ( \begin{array}{c}
    1 \\
    \frac{\bm{\sigma}\cdot \bm{p}_e}{E_e + m_e}
  \end{array}
  \right )
  \chi_{s_e}.
\end{align}

It is noticed the electron wave functions $(G_\kappa, F_\kappa)$
in Ref. \cite{Koshigiri:1979zd} are defined by multiplying
$e^{i\Delta_{\kappa_e}}$ to ours, while
those of Refs. \cite{buhring1963beta,Behrens:1971rbq}
are given by multiplying $\sqrt{2}p_e$ to ours.


The neutrino and antineutrino wave functions are respectively expanded as
\begin{align}
   u_{s_\nu}(\bm{p}_\nu)e^{ i\bm{p}_\nu\cdot\bm{r}}
 & = 
 \sum_{\kappa_\nu,m_\nu,\mu_\nu} \frac{4\pi}{\sqrt{2}} i^{ l_{\kappa_\nu}}  Y_{l_{\kappa_\nu}m_\nu}^*(\hat{p}_\nu)
 (l_{\kappa_\nu} ,m_\nu,1/2,s_\nu|j_{\kappa_\nu},\mu_\nu)
 \left ( \begin{array}{c}
    g_{\kappa_\nu}(r) \chi_{\kappa_\nu}^{\mu_\nu} \\
    i f_{\kappa_\nu}(r)\chi_{-\kappa_\nu}^{\mu_\nu}
  \end{array}
  \right ), \\
  v_{s_\nu}(\bm{p}_\nu)e^{ - i\bm{p}_\nu\cdot\bm{r}}
 & = 
 \sum_{\kappa_\nu,m_\nu,\mu_\nu} \frac{4\pi}{\sqrt{2}} i^{- l_{\kappa_\nu}}  Y_{l_{\kappa_\nu} m_\nu}^*(\hat{p}_\nu)
 (l_{\kappa_\nu} ,m_\nu,1/2,-s_\nu|j_{\kappa_\nu},\mu_\nu)(-1)^{1/2 - s_\nu}\notag\\
 &\times
  \left ( \begin{array}{c}
    - i f_{\kappa_\nu}(r)\chi_{-\kappa_\nu}^{\mu_\nu}\\
    g_{\kappa_\nu}(r) \chi_{\kappa_\nu}^{\mu_\nu} 
  \end{array}
  \right )
\end{align}
with
\begin{align}
  g_\kappa(r) & =  j_{l_\kappa}(p_\nu r), \\
  f_\kappa(r) & =  S_\kappa j_{\bar{l}_\kappa}(p_\nu r),
\end{align}
where $j_l(x)$ is the spherical Bessel function of order $l$,
  $S_\kappa={\rm sgn}(\kappa)$ is the sign of $\kappa$,
and $\bar{l}_\kappa = l_{-\kappa}$.

With the partial wave expansion of the
electron and neutrino wave functions,
one obtains the following form ~\cite{Koshigiri:1979zd}:
\begin{align}
  H_{\mathrm{eff}} =  \frac{G_F V_{ud}}{\sqrt{2}}  \sqrt{\frac{(4\pi)^{3}}{2}}
    \sum_{el}\sum_{neu} \sum_{L,J} (j_e, -\mu_e,j_\nu,\mu_\nu|J,M)(-1)^{1/2-\mu_e}
    \Xi_{JLM}(\kappa_e,\kappa_\nu),
    \label{eq:Heff_beta}
\end{align}
and
\begin{align}
  \Xi_{JLM}(\kappa_e,\kappa_\nu) & =  S_{\kappa_e}\int d\bm{r} \nonumber \\
  &\times  \left\{
    \mp Y_{JM}(\hat{r})V_0(\bm{r})\delta_{L,J}(G_{\kappa_e}(r)g_{\kappa_\nu}(r)S_{0JJ}(\kappa_e,\kappa_\nu)
      + F_{\kappa_e}(r)f_{\kappa_\nu}(r)S_{0JJ}(-\kappa_e,-\kappa_\nu))\right.
        \notag \\
   &     \pm i [Y_L(\hat{r})\otimes\bm{V}(\bm{r})]_{JM}
        (G_{\kappa_e}(r)f_{\kappa_\nu}(r)S_{1LJ}(\kappa_e,-\kappa_\nu)
      - F_{\kappa_e}(r)g_{\kappa_\nu}(r)S_{1LJ}(-\kappa_e,\kappa_\nu))
        \notag \\
 &   +i Y_{JM}(\hat{r})A_0(\bm{r})\delta_{L,J}(G_{\kappa_e}(r)f_{\kappa_\nu}(r)S_{0JJ}(\kappa_e,-\kappa_\nu)
      - F_{\kappa_e}(r)g_{\kappa_\nu}(r)S_{0JJ}(-\kappa_e,\kappa_\nu))
        \notag \\
 &   - \left.[Y_L(\hat{r})\otimes \bm{A}(\bm{r})]_{JM}
        (G_{\kappa_e}(r)g_{\kappa_\nu}(r)S_{1LJ}(\kappa_e,\kappa_\nu)
      + F_{\kappa_e}(r)f_{\kappa_\nu}(r)S_{1LJ}(-\kappa_e,-\kappa_\nu))
      \right\}, \label{eq:xi}
\end{align}
where $[\mathcal{O}_{k_1}\otimes \mathcal{O}_{k_2}^\prime]_{k_3m_3}$
denotes the tensor product.
Here we adopt the following simplified notation for 
$\sum_{el}$ and $\sum_{neu}$:
\begin{align}
  \sum_{el} & = \sum_{\kappa_e,\mu_e}
   i^{-l_{\kappa_e}}e^{i\Delta_{\kappa_e}}(l_{\kappa_e},m_e,1/2,s_e|j_{\kappa_e},\mu_e)
  Y_{l_{\kappa_e}m_e}(\hat{p}_e), 
  \label{eq:sum_electron} \\
  \sum_{neu} & = \sum_{\kappa_\nu,\mu_\nu}
      {} i^{-l_{\kappa_\nu}} (-1)^{1/2 - s_\nu}\notag\\
  &\times [Y_{l_{\kappa_\nu} m_\nu}^*(\hat{p}_\nu)
  (l_{\kappa_\nu},m_\nu,1/2,-s_\nu | j_{\kappa_\nu}, \mu_\nu)
\pm
   Y_{\bar{l}_\kappa m_\nu}^*(\hat{p}_\nu)
(\bar{l}_{\kappa_\nu},m_\nu,1/2,-s_\nu|j_{\kappa_\nu}, \mu_\nu)],
\label{eq:sum_neutrino}
\end{align}
 and 
\begin{align}
  S_{KLJ}(\kappa',\kappa)
  & =
  \sqrt{2(2j_\kappa+1)(2j_{\kappa'}+1)(2l_\kappa+1)(2 l_{\kappa'}+1)(2 K+1)}
 (l_\kappa, 0, l_{\kappa'}, 0|L, 0)
  \left \{ \begin{array}{ccc}
    l_{\kappa'} & 1/2 & j_{\kappa'} \\
    l_\kappa & 1/2 & j_\kappa \\
    L & K & J
  \end{array}
  \right \}.
\end{align}

Using the above form of the effective Hamiltonian, the beta-decay rate is given
by integrating the scattering angles of the neutrino and electron as
\begin{align}
  \Gamma =  \frac{(G_F V_{ud})^2}{\pi^2}
  \int_{m_e}^{E_0} dE_e p_e E_e (E_0 - E_e)^2
  \sum_{J,L,\kappa_e,\kappa_\nu}\frac{1}{2J_i+1}
  |\left<f\right\|\Xi_{JL}(\kappa_e,\kappa_\nu)\left\|i\right>|^2,
  \label{eq:decay_rate}
\end{align}
where $J_i$ is the angular momentum of the initial state.
Neglecting the mass of a neutrino,
the maximum energy of an electron $E_0$
is the
$Q$ value of the nuclear transition,
$Q_\beta$.
See the Appendix \ref{app:decay_rate} for the derivation.

\section{Electron Coulomb Wave function}\label{WF}

\subsection{Parametrization of Lepton Wave Function}

The general formula given in Eq. (\ref{eq:xi}) is ready for the use of
any allowed and forbidden transition rates by evaluating the nuclear
transition density.
However, an explicit formula for the allowed and first-forbidden transitions 
helps extract nuclear structure information from the beta-decay
observables. 
Since the electron Coulomb wave function is rather involved in evaluating the beta-decay rate,
we briefly describe the derivation of the expression of charged lepton wave functions
by iterating the integral equation following  Refs.~\cite{rose1951note,Behrens:1971rbq}.

A Dirac wave function of an electron is given as
\begin{align}
  [\bm{\alpha}\cdot\bm{p}_e + \beta m_e + V_C(r)]\psi_e(\bm{r})
  = E_e \psi_e(\bm{r}).
\end{align}
The electron wave functions $G_\kappa,F_\kappa$ satisfy
the coupled first-order differential equation with the Coulomb potential $V_C$:
\begin{align}
  \frac{dG_\kappa}{dr} + \frac{1+\kappa}{r} G_\kappa
  - (m_e + E_e - V_C) F_\kappa & =  0,
\label{eq-diracg}
  \\
  \frac{dF_\kappa}{dr} + \frac{1-\kappa}{r} F_\kappa
  - (m_e - E_e + V_C) G_\kappa & =  0. 
\label{eq-diracf}
\end{align}
Throughout this paper we keep the electron mass explicit so that in future we can use the formula for the muon neutrino reactions.
Electron wave functions  are parametrized
by taking into account the behavior of the wave function
at the origin $r \sim 0$~\cite{Behrens:1971rbq} as
\begin{align}
  G_{-k}(r) & =  \alpha_{-k} \frac{(p_e r)^{k-1}}{(2k-1)!!}[H_k(r) - h_k(r)],\\
  F_{ k}(r) & =  \alpha_{ k} \frac{(p_e r)^{k-1}}{(2k-1)!!}[H_k(r) + h_k(r)],\\
  G_{ k}(r) & =  \alpha_{k} \frac{(p_e r)^{k-1}}{(2k-1)!!}\frac{r}{R}[D_k(r) + d_k(r)],\\
  F_{-k}(r) & =  -\alpha_{-k} \frac{(p_e r)^{k-1}}{(2k-1)!!}\frac{r}{R}[D_k(r) - d_k(r)].
\end{align}
Here $k>0$ and $H_k(0)=1$ and $h_k(0)=0$.
The normalization of the electron wave functions are determined by
constants $\alpha_\kappa$.
This parametrization of $G_\kappa, F_\kappa$
incorporates the boundary condition of the wave function at the origin.
Then the following set of coupled integral equations
is obtained:
\begin{align}
  H_k(r) & =  1 + \int_0^r \frac{r'}{R}[(-E_e + V_C(r'))D_k(r') + m_e d_k(r')] dr', \\
  h_k(r) & =  \int_0^r \frac{r'}{R}[m_e D_k(r') + (-E_e + V_C(r')) d_k(r')] dr', \\
  \frac{r}{R}D_k(r) & =  \int_0^r \left(\frac{r'}{r}\right)^{2k}
       [(E_e - V_C(r')) H_k(r') + m_e h_k(r')] dr', \\
  \frac{r}{R}d_k(r) & =  \int_0^r \left(\frac{r'}{r}\right)^{2k}
       [m_e H_k(r') + (E_e - V_C(r')) h_k(r')] dr'.
\end{align}
At this stage $R$ is just a parameter of dimension length. We take
$R$ as the nuclear radius though the final
formulas are independent
of the choice of $R$.

\subsection{Iterative Solution of Integral Equation}

Taking into account the boundary condition, $H_k, h_k, D_k$, and $d_k$
are expanded
according to the number of iteration as
\begin{align}
  H_k(r)  &=  1  +  H_k^{(2)}(r) + H_k^{(4)}(r) + \cdots,  \\
  h_k(r)  &=        h_k^{(2)}(r) + h_k^{(4)}(r) + \cdots,  \\
  D_k(r)  &=        D_k^{(1)}(r) + D_k^{(3)}(r) + \cdots,  \\
  d_k(r)  &=        d_k^{(1)}(r) + d_k^{(3)}(r) + \cdots.
\end{align}
The first
iteration of the integral equation gives
\begin{align}
   \frac{r}{R}D_k^{(1)}(r) & =  \int_0^r \left(\frac{r'}{r}\right)^{2k}
       (E_e - V_C(r'))   dr', \\
  \frac{r}{R}d_k^{(1)}(r) & =  \int_0^r \left(\frac{r'}{r}\right)^{2k}
       m_e dr',
\end{align} 
and further iterations give
\begin{align}
  H_k^{(2n)}(r) & =  \int_0^r 
  \left[(-E_e + V_C(r'))
    \frac{r'}{R}  D_k^{(2n-1)}(r') + m_e
    \frac{r'}{R}
    d_k^{(2n-1)}(r')\right]
  dr', \\
  h_k^{(2n)}(r) & =  \int_0^r 
   \left[m_e \frac{r'}{R}D_k^{(2n-1)}(r')
    + (-E_e + V_C(r')) \frac{r'}{R} d_k^{(2n-1)}(r')\right]
    dr',
\end{align}
and
\begin{align}
  \frac{r}{R}D_k^{(2n+1)}(r) & =  \int_0^r \left(\frac{r'}{r}\right)^{2k}
       \left[(E_e - V_C(r')) H_k^{(2n)}(r') + m_e h_k^{(2n)}(r')\right] dr', \\
  \frac{r}{R}d_k^{(2n+1)}(r) & =  \int_0^r \left(\frac{r'}{r}\right)^{2k}
       \left[m_e H_k^{(2n)}(r') + (E_e - V_C(r')) h_k^{(2n)}(r')\right] dr'
\end{align}
for $n=1,2,\dots$.
  The exact electron wave functions
  in terms of $E_e, m_e$, and $V_C$ are obtained from the
  iterative solution of the above equations.

\subsection{LO and NLO electron wave functions}

We denote the leading order (LO) electron wave function  as
\begin{align}
   H_k^{\mathrm{LO}}(r) & =  1, \\
  h_k^{\mathrm{LO}}(r) & =  0, \\
  \frac{r}{R}D_k^{\mathrm{LO}}(r) & =  \frac{r}{R}D_k^{(1)}(r)
  =  \frac{E_e r}{2k+1} + V_{D1}(r), \\
  \frac{r}{R}d_k^{\mathrm{LO}}(r) & =  \frac{r}{R}d_k^{(1)}(r)
  = \frac{m_e r}{2k+1}
\end{align}
 with
\begin{align}
  V_{D1}(r)  & =  - \int_0^r  \left(\frac{r'}{r}\right)^{2k} V_C(r') dr'.
\end{align}

Adding the next-to-leading order (NLO), the NLO wave function is given as
\begin{align}
  H_k^{\mathrm{NLO}}(r)  & =  1 + H_k^{(2)}(r), \\
  h_k^{\mathrm{NLO}}(r)  & =  h_k^{(2)}(r), \\
  \frac{r}{R}D_k^{\mathrm{NLO}}(r)  & =  \frac{r}{R}(D_k^{(1)}(r) + D_k^{(3)}(r)), \\
  \frac{r}{R}d_k^{\mathrm{NLO}}(r)  & =  \frac{r}{R}(d_k^{(1)}(r) + d_k^{(3)}(r)),
  \end{align}
  where
\begin{align}
  H_k^{(2)}(r) & =  - \frac{p_e^2 r^2}{2(2k+1)} + V_{H2}(r),\\
  h_k^{(2)}(r) & =   V_{h2}(r),\\
  \frac{r}{R}D_k^{(3)}(r) & =  - \frac{p_e^2 E_e r^3}{2(2k+1)(2k+3)} + V_{D3}(r), \\
 \frac{r}{R}d_k^{(3)}(r) & =  - \frac{p_e^2 m_e r^3}{2(2k+1)(2k+3)} +  V_{d3}(r)
\end{align}
with
\begin{align}
  V_{H2}(r) & =  \int_0^r
  \left[(V_C(r') - E_e)V_{D1}(r') + E_e\frac{r'}{2k+1}V_C(r')\right]dr',\\
  V_{h2}(r) & =  m_e\int_0^r
  \left[ V_{D1}(r') + \frac{r'}{2k+1} V_C(r')\right]dr',\\
  V_{D3}(r) & =   \int_0^r
  \left(\frac{r'}{r}\right)^{2k}\left[(E_e - V_C(r'))V_{H2}(r') + m_e V_{h2}(r') 
  + \frac{p_e^2 {r'}^2}{2(2k+1)}V_C(r')\right]dr',
\\
V_{d3}(r) & =  
\int_0^r
\left(\frac{r'}{r}\right)^{2k}[m_e V_{H2}(r') + (E_e  - V_C(r')) V_{h2}(r')]dr'.
\end{align}
The explicit expressions of 
$H_k^{(2)},h_k^{(2)},D_k^{(i)}$, and $d_k^{(i)}$
for the uniform charge distribution
are given in the Appendix \ref{app:uniform}.

\section{Decay rate and comparison with the conventional formula}\label{formula}

\subsection{Decay rate}

The beta-decay rate is usually expressed in terms of the Fermi function $F(Z,E_e)$ and
the shape correction factor $C(E_e)$ as \cite{morita1973beta,Schopper:1969jkp}
\begin{align}
  \Gamma & =  \frac{(G_F V_{ud})^2}{2\pi^3}
  \int_{m_e}^{E_0} dE_e p_e E_e (E_0 - E_e)^2 F(Z,E_e) C(E_e). \label{eq:rate1}
\end{align}
Using the matrix element of the effective operator $\Xi_{JLM}$, we obtain
\begin{align}
  F(Z,E_e)C(E_e) & =    \sum_{J,L,\kappa_e,\kappa_\nu}\frac{2\pi}{2J_i+1}
  |\left<f\right\|\Xi_{JL}(\kappa_e,\kappa_\nu)\left\|i\right>|^2, \label{eq:rate2}
\end{align}
and $F(Z,E_e) = \alpha_{-1}^2 + \alpha_1^2$.

\subsection{Allowed and first-forbidden transitions of axial vector current}

We focus on the transition rate due to the space component of the axial vector current.
In the impulse approximation, the axial vector current is given as
\begin{eqnarray}
  \bm{A}(\bm{r}) = g_A\sum_{\tau,\tau^\prime} \sum_{\sigma, \sigma^\prime} 
  \psi^\dagger(\bm{r}\sigma \tau)\psi(\bm{r}\sigma^\prime \tau^\prime)
  \langle \tau|\tau^{\mp}|\tau^\prime\rangle 
  \langle \sigma|\bm{\sigma}|\sigma^\prime\rangle
\end{eqnarray}
with the nucleon field operators $\psi, \psi^\dagger$,
at position $\bm{r}$, spin $\sigma$, and isospin $\tau$.
The transition density $\rho_{JL}(r)$ represented
in the radial coordinate is defined as
\begin{align}
 g_{\rm A} \rho_{JL}(r) &= 
  \left<f\right\| \int d\Omega_r [ Y_L(\hat{r})\otimes \bm{A}(\bm{r})]_{J}
  \left\|i\right>.
\end{align}
The reduced matrix element of the effective operator $\Xi_{JLM}$ is given
in terms of the radial integral of the transition density $\rho_{JL}(r)$
multiplied by combination of the electron and neutrino wave functions with the
coefficients $c_g$ and $c_f$ given in the Appendix~\ref{app:Xi}:
  \begin{align}
\left<f\right\|  \Xi_{JL}(\kappa_e,\kappa_\nu)\left\|i\right>
  & =  \left<f\right\| \int d\bm{r} [Y_L(\hat{r})\otimes \bm{A}(\bm{r})]_{J}[
  c_g G_{\kappa_e}(r)g_{\kappa_\nu}(r)  + c_f F_{\kappa_e}(r)f_{\kappa_\nu}(r) ]\left\|i\right>\notag\\
& = g_{\rm A} \int_0^\infty dr r^2 \rho_{JL}(r)[
  c_g G_{\kappa_e}(r)g_{\kappa_\nu}(r)  + c_f F_{\kappa_e}(r)f_{\kappa_\nu}(r) ]. \label{eq:coef}
  \end{align}

The leading-order formula by Behrens--B\"uhring (LOB) of Ref.~\cite{Behrens:1971rbq} 
conventionally used in the nuclear structure calculations can be derived
by using approximate lepton wave functions in
Eqs.~(\ref{eq:rate2}) and (\ref{eq:coef}). We take the LO electron wave function
and the leading-order approximation of the neutrino wave function.
For the allowed transition
with $\Delta J^\pi=1^+$,
we approximate the $s$-wave wave functions as a constant number:
\begin{align}
  G_{-1}(r) \sim \alpha_{-1},&\quad  g_{-1}(r) \sim 1,\\
  F_{1}(r)  \sim \alpha_1,&\quad    f_1(r) \sim 1,
\end{align}
and neglect all other partial waves.
For the spin-dipole transition
with $\Delta J^\pi=0^-,1^-$, and $2^-$,
in addition to the above approximation
to the $s$-wave wave function, we use the following leading-order approximation
for the $p$-wave wave functions
  \begin{align}
  G_1(r)  \sim \alpha_1 \dfrac{r}{3}[ E_e + m_e + \frac{3V_{D1}(r)}{r}],&\quad
    g_1(r)  \sim \dfrac{p_\nu r}{3}, \\
    F_{-1}(r) \sim - \alpha_{-1}\dfrac{r}{3}[ E_e  - m_e + \frac{3V_{D1}(r)}{r}],&\quad
    f_{-1}(r) \sim - \dfrac{p_\nu r}{3},\\
    G_{-2}(r)  \sim \alpha_{-2} \dfrac{p_e r}{3},&\quad
    g_{-2}(r) \sim \dfrac{p_\nu r}{3},\\
    F_2(r)  \sim \alpha_2 \dfrac{p_e r}{3},&\quad f_2(r) \sim \dfrac{p_\nu r}{3}.
\end{align}

For the allowed $\Delta J^\pi=1^+$ transition, two partial waves  of leptons
$(\kappa_e,\kappa_\nu) = (-1,-1)$ and $(1,1)$ contribute within the LOB,
\begin{align}
  \sum_{\kappa_e,\kappa_\nu} \left|\left<f\right\|\Xi_{J=1,L=0}\left\|i\right>\right|^2
  &\sim  2
 g_{\rm A}^2 \left\{ \left|\int_0^\infty dr r^2  \rho_{10}(r)\left[G_{-1}(r)g_{-1}(r) + \frac{1}{3}F_{-1}(r)f_{-1}(r)\right]\right|^2 \right.
\nonumber \\
&  + \left.\left|\int_0^\infty dr r^2  \rho_{10}(r)\left[\frac{1}{3}G_{1}(r)g_{1}(r) + F_{1}(r)f_{1}(r)\right]\right|^2\right\}
\\
&\sim  2 g_{\rm A}^2(\alpha_{-1}^2 + \alpha_{1}^2)\left|\int_0^\infty dr r^2  \rho_{10}(r)\right|^2.
\end{align}
In the last step, we use
the approximation for the lepton wave functions.
As a result the shape correction factor is given as
\begin{align}
  C(E_e)  =  g_{\rm A}^2\left|\int_0^\infty dr r^2  \sqrt{4\pi}\rho_{10}(r)\right|^2.
\end{align}

The second example is the first-forbidden transition $\Delta J^\pi=0^-$.
The leading-order partial waves are $(\kappa_e,\kappa_\nu) = (-1,1)$
and $(1,-1)$.
We then obtain
\begin{align}
  \sum_{\kappa_e,\kappa_\nu} |\left<f\right\|\Xi_{J=0,L=1}\left\|i\right>|^2
  & \sim  2 g_{\rm A}^2
 \left\{ \left|\int_0^\infty dr r^2  \rho_{01}(r)[G_{-1}(r)g_{1}(r) - F_{-1}(r)f_{1}(r)]\right|^2 \right.
\nonumber \\
&  + \left.\left|\int_0^\infty dr r^2  \rho_{01}(r)[G_{1}(r)g_{-1}(r) - F_{1}(r)f_{-1}(r)]\right|^2\right\}
\\
& \sim \frac{2}{9} g_{\rm A}^2
\left\{\alpha_{-1}^2\left|\int_0^\infty dr r^3  \rho_{01}(r)
\left[ p_\nu +E_e - m_e + 3 \frac{V_{D1}(r)}{r}\right]\right|^2 \right.
\nonumber \\
& +\left.\alpha_{1}^2\left|\int_0^\infty dr r^3
\rho_{01}(r)\left[ p_\nu +E_e + m_e + 3 \frac{V_{D1}(r)}{r}\right]\right|^2\right\}.
\end{align}
In order to compare our formula with LOB,
for example,  Eq. (10.56) of Ref.~\cite{sch66},
introducing nuclear matrix elements
\begin{align}
  \omega & =  g_{\rm A}\sqrt{4\pi}\int_0^\infty dr r^3 \rho_{01}(r), \\
  \xi\omega' & =  g_{\rm A}\sqrt{4\pi}\int_0^\infty dr r^2 \rho_{01}(r) V_{D1}(r),
\end{align}
we obtain
\begin{align}
  C(E_e) = \zeta_0^2 + \frac{\omega^2m_e^2}{9}
  - \frac{2}{3}\frac{\mu_1\gamma_1 m_e^2}{E_e} \zeta_0 \omega,
\end{align}
where
\begin{align}
  \zeta_0 & =  \frac{E_0 \omega}{3} + \xi\omega',\\
  \gamma_k & =  \sqrt{k^2- (\alpha Z)^2}, \\
  \mu_k & = \frac{k}{\gamma_k}\frac{E_e}{m_e}
  \frac{\alpha_{-k}^2 - \alpha_k^2}{\alpha_{-1}^2 + \alpha_1^2}.
\end{align}  
Here $\alpha$ is the fine structure constant.
A similar comparison can be done for the transitions to $1^-$ and $2^-$ states,
and we can confirm the use of the approximate lepton wave function within
our formalism leads to the `conventional' formula of the decay rate.

\section{Analysis with a schematic model}\label{model}

In the following, we examine
the validity of the approximation for the electron wave function
proposed in this work by using a schematic model of transition
density. Three sets of treatment of the lepton
wave function (i) exact, (ii) LO, and (iii) NLO are defined.
By (i) `exact', we use the electron wave function obtained by a
numerical solution of the Dirac equation and the spherical
Bessel function for the neutrino wave function.
In (ii) LO and (iii) NLO, we approximate the electron wave function by
the LO and NLO wave functions described in the previous section.
Notice that we do not expand the neutrino wave function.
We use the uniform charge distribution for the
nuclear charge with a charge radius $R_A=1.2\times A^{1/3}$ fm, 
and a transition density  given by a sum of two Gaussians. 
The analytic expressions for the LO and NLO terms of the
electron wave functions are summarized in the Appendix \ref{app:uniform}.
We found that the numerical results of LO are very close to those
of `conventional' formula LOB.

The explicit forms
of the  LO and NLO approximation of the electron wave functions
of $\kappa_e=-1$ with
the $s$-wave large component ($G_{-1}$) and
the $p$-wave small component ($F_{-1}$) are given as
\begin{align}
  G_{-1}^{\rm LO}(r) = & \alpha_{-1}, \\
  F_{-1}^{\rm LO}(r) = & -\alpha_{-1}  r\left[ \frac{E_e-m_e}{3} +\xi s_1(x)\right],\\
  G_{-1}^{\rm NLO}(r) = & G_{-1}^{\rm LO}(r) + \alpha_{-1}r^2\left[- \frac{p_e^2}{6} + \xi (E_e s_2(x) - m_e h_2(x)) + \xi^2 t_2(x)\right], \\
  F_{-1}^{\rm NLO}(r) = & F_{-1}^{\rm LO}(r)  -\alpha_{-1}r^3\left[- \frac{p_e^2 (E_e -m_e)}{30}+\xi(p_e^2  s_3(x) + m_e(m_e - E_e) t_3(x)) \right.  \notag \\
      & \left. + \xi^2(E_e w_3(x) - m_e z_3(x))+ \xi^3 y_3(x)\right],
\end{align}
where $x=r/R_A$ and $\xi = \alpha Z/(2R_A)$.

\begin{figure}[t]
\begin{center}
\includegraphics[width=0.45\textwidth]{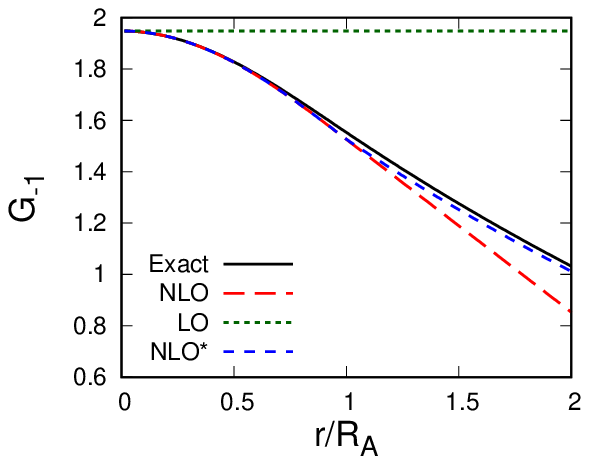}
\includegraphics[width=0.45\textwidth]{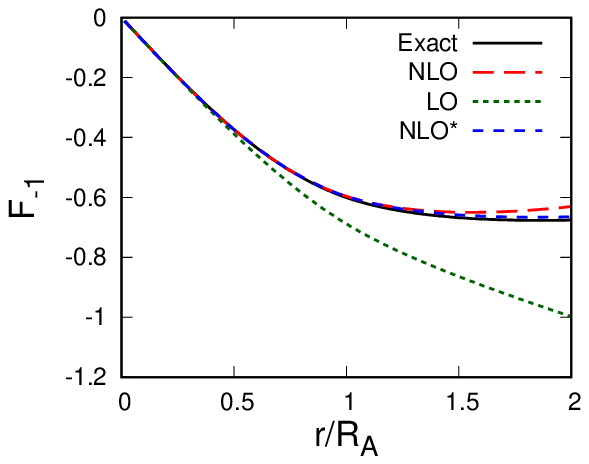}

\caption{\label{fig-ewf1} Electron wave functions $G_{-1}$ (left) and $F_{-1}$ (right) 
  for $Z=82, A=208$, $\kappa_e=-1$ and $E_e=10$ MeV.
The NLO wave function of this work
and the LO wave function approximately corresponding to `conventional'
(LOB) 
    are compared.
  The NLO$^*$ denotes the one
  obtained by connecting
  the NLO wave function with the point Coulomb regular and irregular
  wave functions at $r=R_A$.
}
\end{center}
\end{figure}

\begin{figure}
\centering
\includegraphics[width=0.45\textwidth]{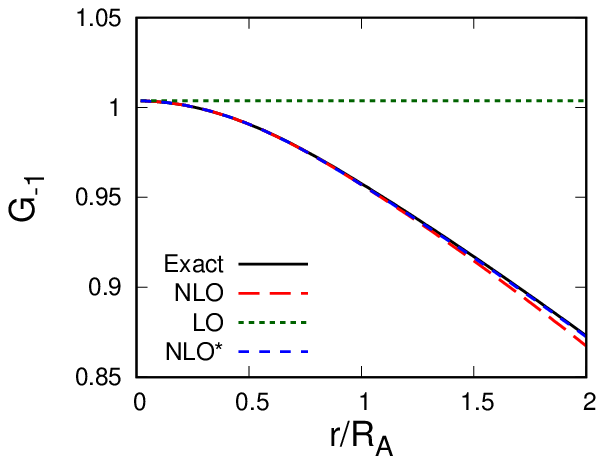}
\includegraphics[width=0.45\textwidth]{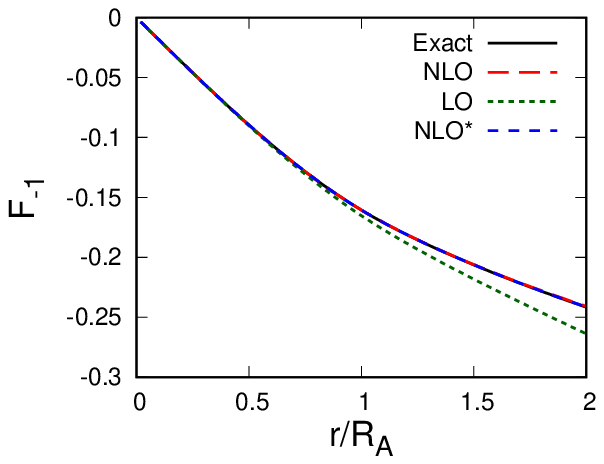}

\caption{Same as Fig.~\ref{fig-ewf1} but for $Z=28$ and $A=80$.}
\label{fig-ewf2}
\end{figure}

Figure~\ref{fig-ewf1} shows the electron wave functions $G_{-1}$ and $F_{-1}$
at $E_e=10$ MeV for $Z=82$ and $A=208$.
The `exact' and the `LO' wave functions 
are shown by the solid and short-dashed curves, respectively. The deviation of the
LO wave function from the exact one grows as $r$ increases.
One can see that the deviation is larger for an $s$-wave than a $p$-wave wave function.
By taking into account the NLO correction,
the wave functions are greatly improved, but a 
slight deviation from the `exact' wave function still remains for a larger $r$ region.
For the uniform charge distribution,
the Coulomb potential for $r>R_A$ agrees with the point Coulomb potential.
Therefore, by connecting the NLO wave function with the
combination of the analytic form of the 
regular and irregular point Coulomb wave functions, 
we can obtain the electron
wave functions for $r>R_A$ with improved accuracy (NLO$^*$).
This is indeed the case, as shown in the blue dashed curves in Fig.~\ref{fig-ewf1}.
Figure~\ref{fig-ewf2} is the same as Fig.~\ref{fig-ewf1} but 
for $Z=28, A=80$.  
One sees that the effects of the NLO correction are smaller than
in the $Z=82$ case, 
though the deviation of the LO wave function is distinct for the $s$-wave.

For $\beta^+$ decay, the sign changes for the odd power terms of $\xi$.
The Coulomb potential 
enters in the Dirac equation
in the form of  $E_e-V_C$ as given in
Eqs.~(\ref{eq-diracg}) and (\ref{eq-diracf}).
The Coulomb effect is constructive to $E_e$
for an electron, while it is destructive for a positron.
For $E_e > |V_C|$, the deviation from the LO wave
function becomes smaller for a positron than for an electron.

\begin{figure}
\centering
\includegraphics[width=0.45\textwidth]{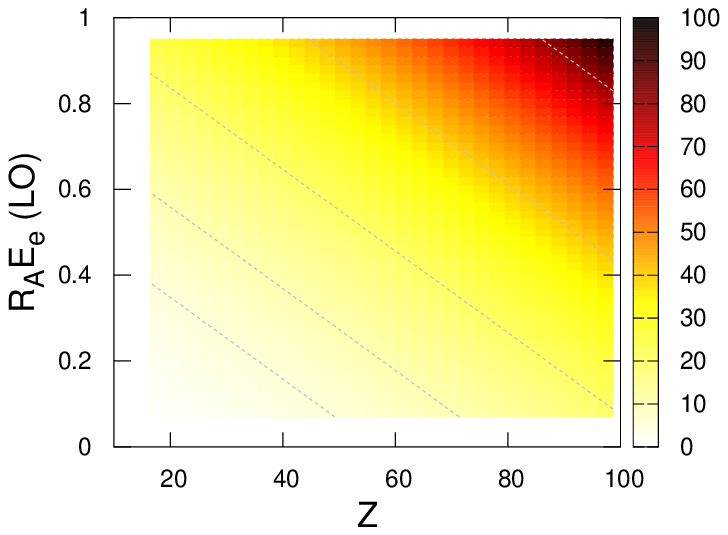}
\includegraphics[width=0.45\textwidth]{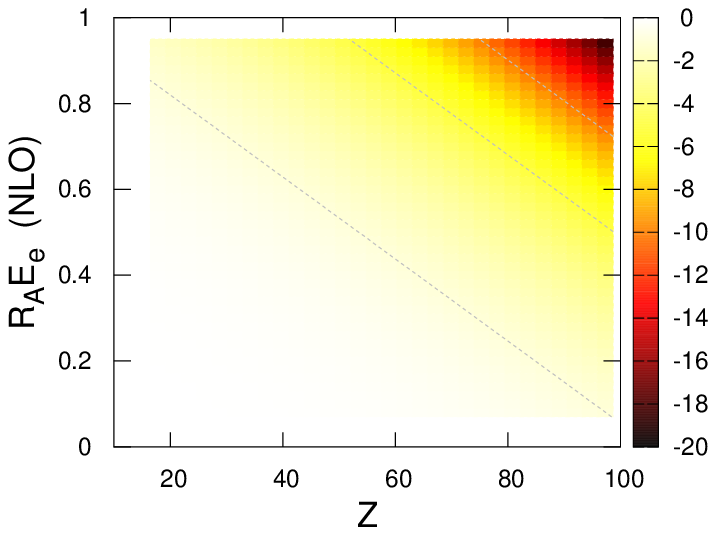}

\caption{Relative deviation $[G_{-1}({\rm approx.})/G_{-1}({\rm exact})-1]\times 100$ as a function of $R_AE_e$ and $Z$.
  The LO wave function (left) and NLO wave function (right) is used for the
  approximate electron wave function.}
\label{fig-ewf3}
\end{figure}

The $E_e$ and $Z$ dependence of the NLO correction is
parametrized essentially by two
non-dimensional parameters $R_A E_e$ and $R_A \xi = \alpha Z/2$.
Figure~\ref{fig-ewf3} shows the deviation of the approximate LO and NLO
electron wave functions $G_{-1}$ at the nuclear surface
$r=R_A$, $[G_{-1}({\rm approx.})/G_{-1}({\rm exact}) -1] \times 100$,
as a function of $R_A E_e$ and $Z$.
One sees a considerable deviation  for a larger nuclear charge $Z$ and a
higher 
$R_A E_e$ value. The LO approximation overestimates the amplitude of the wave function at the
nuclear surface.  
By including the NLO correction, an error is notably reduced.
For $E_e=10$ MeV, 
the case of $Z=82, A=208$ and $Z=28, A=80$ corresponds to
$R_A E_e \sim 0.30,  R_A \xi \sim 0.36$  and 
 $R_A E_e \sim 0.26,  R_A \xi \sim 0.10$, respectively.

\begin{figure}
  \begin{center}
    \centering
    \includegraphics[width=0.6\textwidth]{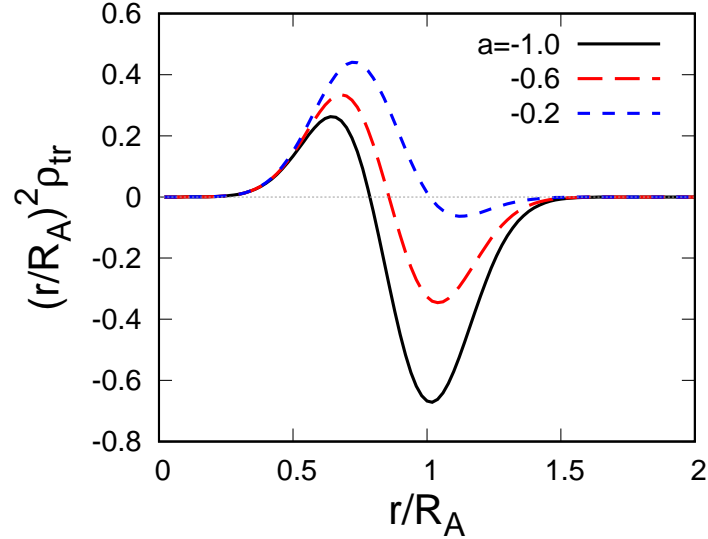}
  \end{center}

\caption{Transition density used in the present simple model analysis
  multiplied by $(r/R_A)^2$ with ${\cal N}=1$
  as a function of the scaled radial coordinate $r/R_A$.
  See text for details.
  The dotted horizontal line indicates zero.}
  \label{fig-trd1}
\end{figure}

The difference between the LO and `exact' lepton
wave functions observed above certainly affects the beta-decay rate. 
The magnitude of the effect depends on 
the transition density of nuclear weak currents.
To examine the effects on the beta-decay rate,
we take the following simple form of the transition density
for the Gamow--Teller and spin-dipole transitions:
\begin{align}
  \rho_{\rm tr} = {\cal N}[a\  e^{- (r - r_1)^2/b^2} + e^{- (r - r_2)^2/b^2}]. \label{eq:tra_den_model}
\end{align}
Here we take $r_1=0.9R_A, b=R_A/4$ and $r_2=3 r_1/4$.
By varying  $-1 \le a \le -0.2$,
we investigate 
the validity of the approximation for the
electron wave function on the decay rate.
The transition density multiplied by $r^2$, $(r/R_A)^2\rho_{\rm tr}(r)$, is shown in
Fig.~\ref{fig-trd1}
  with $a$ as a parameter.
  For $a= -0.2$, the contribution of the $r<R_A$ region
  is predominant for the   transition matrix element,
while for $a= -1$, the $r \sim R_A$ region gives a prevailing contribution to the matrix element.
For $a=-0.6$, a strong suppression of the matrix element would take place.

\begin{figure}
  \centering
  \includegraphics[width=0.45\textwidth]{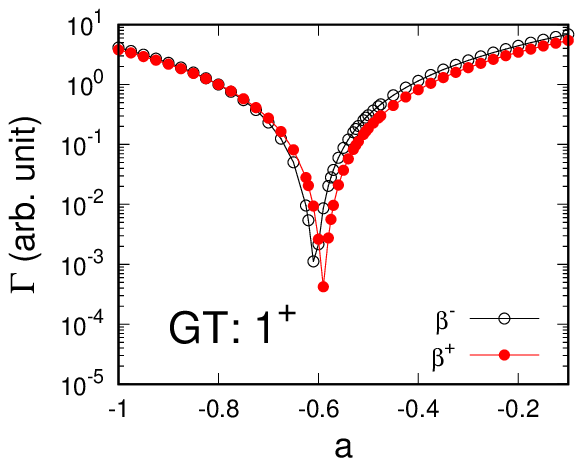}
  \includegraphics[width=0.45\textwidth]{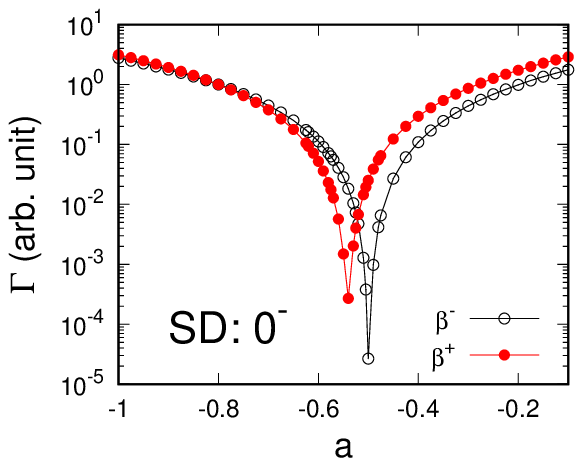}

  \caption{$a$ dependence of the decay rate
    for the Gamow--Teller (GT) $\Delta J^\pi=1^+$ (left) and spin-dipole
    (SD) $\Delta J^\pi=0^-$ (right) transitions
    of $\beta^\pm$ decay.
    Decay rates are calculated with the `exact' wave function for $Z=82 \pm 1$,
    $A=208$, $E_0=10$ MeV and normalized as unity at $a=-0.8$. 
    The lines are guide for the eye.}
  \label{fig-rate1}
\end{figure}

  In what follows, we investigate the decay rate using the transition density (\ref{eq:tra_den_model}). First, 
  we use the exact electron wave function and set $Z=82 \pm 1$ for $\beta^\mp$
  decay and $A=208$.
  The decay rate 
  shown in Fig.~\ref{fig-rate1} with open symbols for $\beta^-$ decay 
 is given in arbitrary unit normalized to unity at $a=-0.8$.
 The lines in the figure are guide for the eye.
 Strong suppression of the transition rate of the allowed Gamow--Teller
 transition is seen around $a=-0.6$,
while it happens around $a = -0.5$ for the $0^-$ transition.
For the first-forbidden transition, an extra factor $r$ of the operator
moves slightly the minimum position of the matrix element. 
We obtain a similar $a$ dependence of the $\beta^+$ decay rate for the normalized rate as shown by filled symbols.

\begin{figure}
  \centering
  \includegraphics[width=0.45\textwidth]{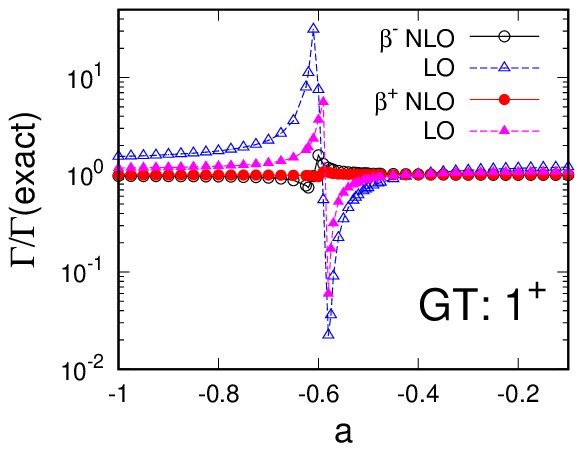}
  \includegraphics[width=0.45\textwidth]{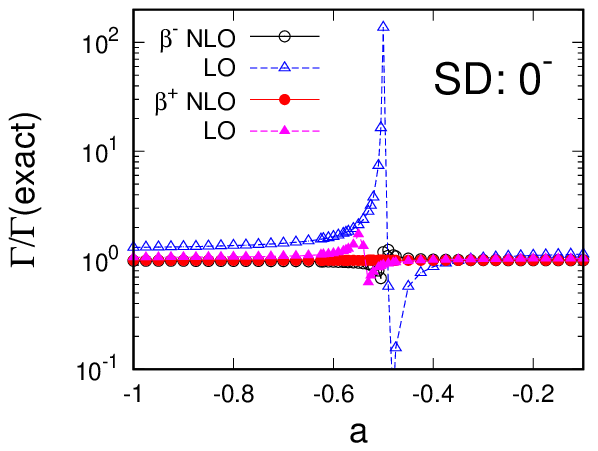}

  \caption{Ratio of the transition rate for the Gamow--Teller (GT) $\Delta J^\pi=1^+$ (left)
    and spin-dipole (SD) $\Delta J^\pi=0^-$ (right) transitions.
    The NLO $\Gamma({\rm NLO})/\Gamma({\rm exact})$
      and the LO $\Gamma({\rm LO})/\Gamma({\rm exact})$
      for $\beta^{\pm}$ decay are compared.
      The decay rate is calculated for $Z=82 \pm 1$,  $A=208$, and $E_0=10$ MeV.
          The lines are guide for the eye.
  }
  
  \label{fig-rate2}
\end{figure}

We then study the validity of the approximation of the electron wave function
thoroughly.
Figure~\ref{fig-rate2} shows the ratio of the $\beta^-$ decay rate 
calculated with the LO wave function to that with the exact one, 
$\Gamma({\rm LO})/\Gamma({\rm exact})$, 
drawn by the  dashed curve with open symbols
(filled symbols for $\beta^+$ decay), 
and compares it
with the ratio of the rate calculated with the NLO wave function to that with the exact one, 
$\Gamma({\rm NLO})/\Gamma({\rm exact})$, depicted by the solid 
curve with open symbols 
(filled symbols for $\beta^+$ decay).
A large deviation is suspected to occur with the use of the approximate lepton wave functions
in particular when a delicate cancellation of the radial integral takes place.
The deviation from the exact calculation 
is large for $a < -0.6$, where the contribution at 
$r \sim R_A$ is more important than $r < R_A$.
The use of the augmented  NLO electron wave function
significantly improves for $a < -0.7$, and works reasonably well
even when a severe cancellation between the inner and the 
outer contribution of the integration
takes place around $a = -0.6$.

\begin{figure}
  \centering
  \includegraphics[width=0.45\textwidth]{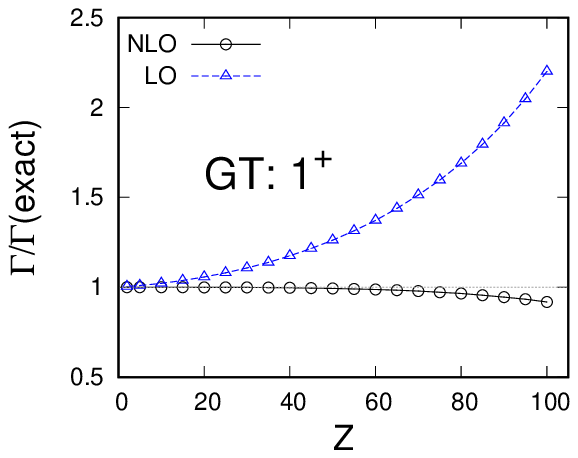} 
  \includegraphics[width=0.45\textwidth]{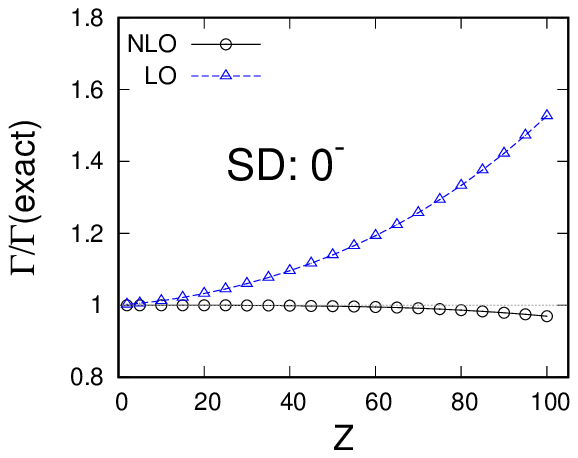} \\
  \includegraphics[width=0.45\textwidth]{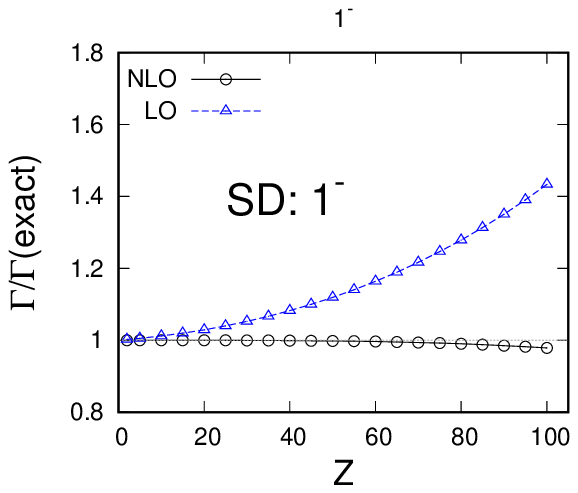}
  \includegraphics[width=0.45\textwidth]{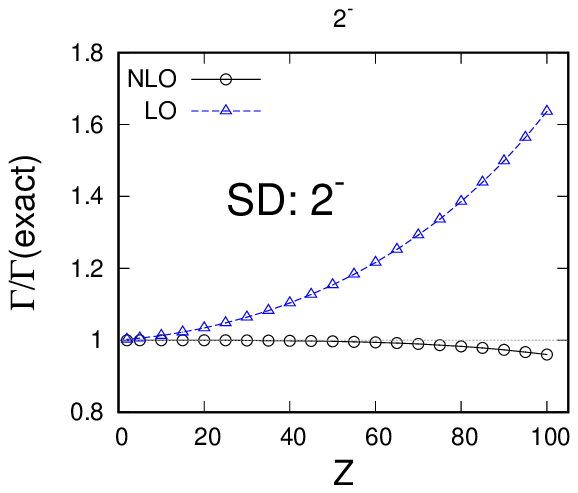}

    \caption{Ratio of the $\beta^-$ decay rate,
        $\Gamma({\rm NLO})/\Gamma({\rm exact})$ and
        $\Gamma({\rm LO})/\Gamma({\rm exact})$
    for the Gamow--Teller (GT) $\Delta J^\pi=1^+$,
    and the spin-dipole (SD) $\Delta J^\pi=0^-$, $1^-$ and $2^-$ transitions.
    The decay rate is calculated for $E_0=10$ MeV and $A=2Z$.
      The dotted horizontal line indicates unity.}
  \label{fig-zdep}
\end{figure}

At the end of the study with the schematic model, we investigate
the $Z$ dependence of the NLO correction. 
Figure~\ref{fig-zdep} shows
the $Z$ dependence of the $\beta^-$ decay rate for $E_0=10$ MeV 
calculated by the LO, $\Gamma({\rm LO})/\Gamma({\rm exact})$, 
and by the NLO, $\Gamma({\rm NLO})/\Gamma({\rm exact})$, 
for the Gamow--Teller and spin-dipole transitions. 
Here the transition density with a moderate cancellation of the matrix element with $a=-0.8$ is used. 
One sees a simple use of the LO or conventional (LOB)
formula overestimates the exact rate by about 50--100\% for heavy nuclei. 
  This suggests that
  the $B(GT)$ value extracted from the beta-decay rate using the LO
  can be underestimated for the transition involving heavy neutron-rich nuclei.
However, it is apparent that 
our NLO approximation works well for a wide range of the nuclear charge. 

\section{EDF transition density and NLO electron wave function}\label{EDF}

To investigate the validity of our formalism in realistic cases, 
we use the transition densities
microscopically calculated by a nuclear energy-density functional (EDF) method. 
Since the details of the formalism can be found in Ref.~\cite{yos13}, 
here we recapitulate the basic equations relevant to the present study.  
In the framework of the nuclear EDF method we employ, 
the ground state of a mother nucleus is described by solving the 
Kohn--Sham--Bogoliubov (KSB) equation~\cite{dob84}
\begin{align}
\begin{bmatrix}
h^{q}(\boldsymbol{r} \sigma)-\lambda^{q} & \tilde{h}^{q}(\boldsymbol{r} \sigma) \\
\tilde{h}^{q}(\boldsymbol{r} \sigma) & -h^{q}(\boldsymbol{r} \sigma)+\lambda^{q}
\end{bmatrix}
\begin{bmatrix}
\varphi^{q}_{1,\alpha}(\boldsymbol{r} \sigma) \\
\varphi^{q}_{2,\alpha}(\boldsymbol{r} \sigma)
\end{bmatrix} 
= E_{\alpha}
\begin{bmatrix}
\varphi^{q}_{1,\alpha}(\boldsymbol{r} \sigma) \\
\varphi^{q}_{2,\alpha}(\boldsymbol{r} \sigma)
\end{bmatrix}, \label{HFB_eq}
\end{align}
where 
the KS 
potentials $h$ and $\tilde{h}$ are given by the EDF. 
An explicit expression of the potentials can be found for example in the Appendix of Ref.~\cite{kas21}. 
The chemical potential $\lambda$ is determined so as to give the desired nucleon number as an average value.
The superscript $q$ denotes 
n (neutron, $\tau_z= 1$) or p (proton, $\tau_z =-1$).

The excited states $| f; J^\pi \rangle$ in a daughter nucleus are described as 
one-phonon excitations built on the ground state $|i\rangle$ of the mother nucleus as 
\begin{align}
| f; J^\pi \rangle &= \Gamma^\dagger_{f} |i \rangle, \\
\Gamma^\dagger_{f} &= \sum_{\alpha \beta}\left\{
X_{\alpha \beta}^f a^\dagger_{\alpha,{\rm n}}a^\dagger_{\beta, {\rm p}}
-Y_{\alpha \beta}^f a_{\beta,{\rm p}}a_{\alpha,{\rm n}}\right\},
\end{align}
where $a^\dagger_{\rm n} (a^\dagger_{\rm p})$ and $a_{\rm n} (a_{\rm p})$ are 
the neutron (proton) quasiparticle (labeled by $\alpha$ and $\beta$) creation and annihilation operators that 
are defined in terms of the solutions of the KSB equation (\ref{HFB_eq}) with the Bogoliubov transformation. 
The phonon states, the amplitudes $X^f, Y^f$ and the vibrational frequency $\omega_f$, 
are obtained in the proton--neutron quasiparticle-random-phase approximation (pnQRPA). 
The residual interactions entering into the pnQRPA equation 
are given by the EDF self-consistently. 
With the solutions of the pnQRPA equation, the transition density is given as
\begin{align}
 g_A \delta \bm{\rho}_{f; J^\pi}(\boldsymbol{r})=\langle f; J^\pi| \bm{A}(\bm{r})  |i\rangle 
=  \langle i|[\Gamma_{f}, \bm{A}(\bm{r})]|i\rangle
\end{align}
in a standard quasi-boson approximation. 
One obtains the transition density in the radial coordinate as
\begin{align}
\rho_{JLK}(r)=\int d\Omega_r [ Y_L(\hat{r})\otimes \delta\bm{\rho}(\bm{r})]_{JK},
\end{align}
which is independent of $K$ in the present case for spherical systems. 
Thus, the input transition density is obtained by $\rho_{JL}(r)=\sqrt{2J+1}\rho_{JL0}(r)$.

We apply our formula for the medium-heavy Ni and Sn isotopes. 
Since a considerable contribution of the first-forbidden transition is predicted in the Sn isotopes~\cite{mus16}, 
we take ${}^{160}$Sn as an example in the present study. 
Furthermore, 
an interplay between the allowed and first-forbidden transitions has been discussed around ${}^{78}$Ni~\cite{yos19}, 
and we thus take ${}^{80}$Ni as a target of the present study as well and employ the same Skyrme and pairing EDF as in Ref.~\cite{yos19}. 
Within the pnQRPA, the maximum electron energy is given  as 
$E_0 = B(Z+1,N-1)-B(Z,N)+(m_{\rm n}-m_{\rm p}-m_e)
\simeq \lambda^{\rm n}-\lambda^{\rm p}-\omega+0.78$ MeV for $\beta^-$ decay~\cite{eng99}.

\begin{figure}
\centering
  \includegraphics[width=0.45\textwidth]{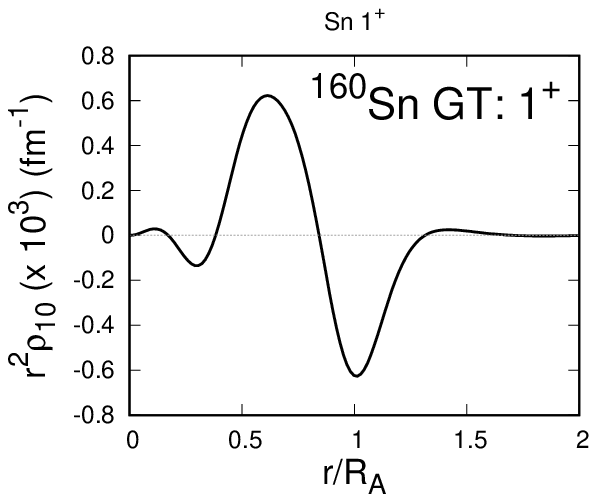}
  \includegraphics[width=0.45\textwidth]{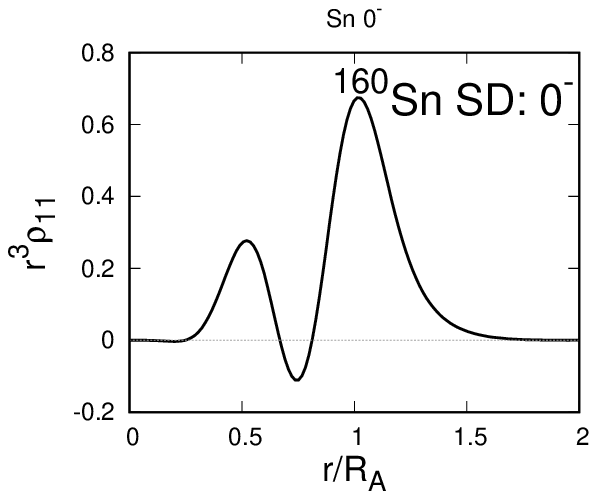}

  \caption{
    Transition density $r^{2+L}\rho_{JL}$ of
     (left) Gamow--Teller transition
    $J^\pi=0^+ \rightarrow 1^+$ ($E_0=12.1$ MeV)
    multiplied by $10^{3}$ 
    and (right) spin-dipole transition
    $J^\pi=0^+ \rightarrow 0^-$ ($E_0=16.1$ MeV) of ${}^{160}$Sn.
  The dotted horizontal line indicates zero.}
  \label{fig-trdns-sn}
\end{figure}

The transition densities $\rho_{JL}$ of the $J^\pi=1^+$ ($E_0 =12.1$ MeV)
  and $0^-$ (16.1 MeV) states in $^{160}$Sn are
  shown in Fig.~\ref{fig-trdns-sn}. Those states give the largest contribution 
  to the transition rate for each $J^\pi$.
One sees there are nodes similarly to the transition densities of the schematic model.
In such a case, the contribution around the nuclear 
surface $r \sim R_A$ is important. 

Using the transition densities microscopically calculated  by the EDF method, we evaluate 
the half-life of $\beta^-$ decay of the allowed Gamow--Teller and
the first-forbidden spin-dipole transitions 
of ${}^{80}$Ni and ${}^{160}$Sn.  The $\beta$ decay rates are calculated 
within the impulse approximation for the space component of the axial vector current only. Here we use
  the effective axial vector coupling constant $g_{\rm A} = 1$.
The  half-life is calculated using the `exact' formula without approximation
for the lepton wave functions. 
The contribution of all the states up to $E_0 \sim 14$ MeV (16 MeV) for ${}^{80}$Ni (${}^{160}$Sn) 
are included. 
Shown in Tab.~\ref{tbl-hfl-nisn} is the half-life thus calculated for each $J^\pi$.
We show the ratios of the half-life 
$t_{1/2}$(LO)/$t_{1/2}$(exact) and 
$t_{1/2}$(NLO)/$t_{1/2}$(exact) in the table as well.
As suspected the LO overestimates the transition rate by about 5 to 15\% depending
on the type of the transition and nuclide. Therefore, the half-lives are underestimated. 
The deviation from the `exact'
calculation is larger for Sn than for Ni.
Introducing the NLO correction, those errors are nicely restored, as shown
in the third column of Tab.~\ref{tbl-hfl-nisn}. 
We can therefore argue that
our simple NLO formula is  very effective in realistic calculations.

\begin{table}
\centering
\caption{Half life $t_{1/2}$
  of the $\beta^-$ decay of $^{80}$Ni and $^{160}$Sn
to the daughter nucleus with the states $J^\pi$.
The ratios of the half life 
$t_{1/2}$(LO)/$t_{1/2}$(exact) and 
$t_{1/2}$(NLO)/$t_{1/2}$(exact) are shown in the second and the third column.}
\begin{tabular}{cccccccc}
\hline\hline
  & \multicolumn{3}{c}{${}^{80}$Ni} &&   \multicolumn{3}{c}{$^{160}$Sn} \\ \cline{1-4} \cline{6-8}

$J^\pi$ & $t_{1/2} ({\rm s})$ & LO & NLO && $t_{1/2} ({\rm s})$ & LO & NLO \\ \cline{1-4} \cline{6-8}  
  $1^+$ & 3.50 $\times 10^{-1}$ & 0.962 & 1.00 && 2.39$\times 10^{-3}$ & 0.874  & 1.00 \\
  $0^-$ & 1.08 & 0.928 & 1.00 && 1.34$\times 10^{-2}$ & 0.874  & 1.00 \\
  $1^-$ & 3.02 & 0.943 & 1.00 && 5.18$\times 10^{-2}$ & 0.895  & 1.00 \\
  $2^-$ & 2.18 & 0.942 & 1.00 && 1.42$\times 10^{-1}$ & 0.857  & 1.00 \\
\hline\hline
\end{tabular}
\label{tbl-hfl-nisn}
\end{table}

\section{Summary}\label{summary}

We have investigated the Coulomb effects on the beta-decay rate. 
The decay rate is determined by the product of the lepton and hadron current densities. 
A widely used formula relies on the fact that the low-energy lepton wave functions in a nucleus 
can be well approximated  
by a constant and linear to the radius for the $s$-wave and $p$-wave wave functions, respectively. 
We found, however, 
the Coulomb wave function is conspicuously different from such a 
simple approximation for heavy nuclei with large $Z$
by numerically solving the Dirac equation. 
We then have proposed formulas of the nuclear beta-decay rate that are useful in a practical calculation. 

In our proposed formulas, 
the neutrino wave function is treated exactly as a plane wave, while the electron wave function is 
obtained by iteratively solving the integral equation; thus, we can control the uncertainty of the approximate electron wave function order by order. 
The leading-order approximation gives a formula that is almost equivalent to the widely used one and overestimates the decay rate 
by about 50--100\% for heavy nuclei with $Z \sim 80$. 
We demonstrated that the next-to-leading-order formula reproduces well the exact result for a 
schematic transition density as well as a microscopic one obtained by a nuclear energy-density functional method. 
For the beta decay involving heavy neutron-rich nuclei, the NLO will be needed  
for the determination of the Gamow--Teller strength 
from the beta-decay rate.

We considered only the space component of the axial vector currents and kept only the lowest multipoles. 
The time components as well as the vector currents can have a comparable contribution to the decay rate, 
and we plan to present these improvements in a sequel to the present article. 
The beta decay provides a unique spectroscopic tool of exotic nuclei,
that is, the angular correlation 
contains rich information of nuclear structure. 
Furthermore, the electron/muon capture is an important process in the application 
to astrophysics and fundamental physics. 
It is straightforward to extend our formalism towards these directions. 

\section*{Acknowledgment}
We would like to thank Prof. K. Koshigiri for useful discussions.
This work was in part supported by the JSPS KAKENHI
Grants Nos. JP18H01210,  JP18H04569,  JP18K03635,  JP19H05104,
JP19H05140, and  JP19K03824, 
 the Collaborative Research Program 2019--2020, Information Initiative Center, Hokkaido University, 
and the JSPS/NRF/NSFC A3 Foresight Program ``Nuclear Physics in the 21st Century.'' 
The nuclear EDF calculation was performed on CRAY XC40 
at the Yukawa Institute for Theoretical Physics, Kyoto University.

\appendix

\section{Derivation of the decay rate \label{app:decay_rate}}

In this appendix, we show the derivation of the decay rate Eq.~\eqref{eq:decay_rate} for $\beta^-$ decay.
In the present case, it is useful to
expand the effective Hamiltonian in terms of the angular momentum.
With the partial wave expansion of the neutrino wave function, we get
\begin{align}
\left(1-\gamma_5\right)v^{s_\nu}\left(p_\nu\right)e^{-i\bm{p}_\nu\cdot\bm{r}}=& 4\pi\sum_{neu}
\begin{pmatrix}
-if_{\kappa_\nu}(r)\chi_{-\kappa_\nu}^{\mu_\nu} \\
g_{\kappa_\nu}(r)\chi_{\kappa_\nu}^{\mu_\nu}
\end{pmatrix},
\end{align}
where we use
the abbreviated notation (upper sign) defined
in Eq.~\eqref{eq:sum_neutrino}.
Since the neutrino mass is negligible, we used here $g_{\kappa}(r)=-S_\kappa f_{-\kappa}(r)$ and $f_{\kappa}(r)=S_\kappa g_{-\kappa}(r)$.
We also have an alternative expression: 
\begin{align}
\left(1-\gamma_5\right)v^{s_\nu}\left(p_\nu\right)e^{-i\bm{p}_\nu\cdot\bm{r}}=& 4\pi\sum_{neu}
\begin{pmatrix}
-g_{\kappa_\nu}(r)\chi_{-\kappa_\nu}^{\mu_\nu} \\
if_{\kappa_\nu}(r)\chi_{\kappa_\nu}^{\mu_\nu}
\end{pmatrix}.
\end{align}
Thus, we obtain two equivalent expressions:
\begin{align}
&\psi^{s_e(-)}_{p_e}\left(\bm{r}\right)\gamma^\nu\left(1-\gamma_5\right)v^{s_\nu}\left(p_\nu\right)e^{-i\bm{p}_\nu\cdot\bm{r}} \nonumber\\
=&\left(4\pi\right)^2\sum_{el}\sum_{neu}\left(G_{\kappa_e}\left(r\right)\chi_{\kappa_e}^{\mu_e\dagger},iF_{\kappa_e}\left(r\right)\chi_{-\kappa_e}^{\mu_e\dagger}\left(\hat{r}\right)\right)\gamma^\nu
\begin{pmatrix}
-if_{\kappa_\nu}(r)\chi_{-\kappa_\nu}^{\mu_\nu} \\
g_{\kappa_\nu}(r)\chi_{\kappa_\nu}^{\mu_\nu}
\end{pmatrix} \\
=&\left(4\pi\right)^2\sum_{el}\sum_{neu}\left(G_{\kappa_e}\left(r\right)\chi_{\kappa_e}^{\mu_e\dagger},iF_{\kappa_e}\left(r\right)\chi_{-\kappa_e}^{\mu_e\dagger}\left(\hat{r}\right)\right)\gamma^\nu
\begin{pmatrix}
-g_{\kappa_\nu}(r)\chi_{-\kappa_\nu}^{\mu_\nu} \\
if_{\kappa_\nu}(r)\chi_{\kappa_\nu}^{\mu_\nu}
\end{pmatrix},
\end{align}
where we used Eq.~\eqref{eq:sum_electron}.

As in Eq.~\eqref{eq:hadron_current}, the hadron current is composed of
the vector and axial vector components.
We can thus write
\begin{align}
&\psi^{s_e(-)}_{p_e}\left(\bm{r}\right)\gamma^\nu\left(1-\gamma_5\right)v^{s_\nu}\left(p_\nu\right)e^{-i\bm{p}_\nu\cdot\bm{r}}J_\nu \nonumber\\
=&\left(4\pi\right)^2\sum_{el}\sum_{neu}\left(G_{\kappa_e}\left(r\right)\chi_{\kappa_e}^{\mu_e\dagger}\left(\hat{r}\right),iF_{\kappa_e}\left(r\right)\chi_{-\kappa_e}^{\mu_e\dagger}\left(\hat{r}\right)\right) \nonumber\\
&\times\left\{\left(\gamma^0 V_0-\bm{\gamma}\cdot\bm{V}\right)
\begin{pmatrix}
-g_{\kappa_\nu}(r)\chi_{\kappa_\nu}^{\mu_\nu}\left(\hat{r}\right) \\
if_{\kappa_\nu}(r)\chi_{-\kappa_\nu}^{\mu_\nu}\left(\hat{r}\right)
\end{pmatrix}-\left(\gamma^0A_0-\bm{\gamma}\cdot\bm{A}\right)
\begin{pmatrix}
-if_{\kappa_\nu}(r)\chi_{-\kappa_\nu}^{\mu_\nu} \\
g_{\kappa_\nu}(r)\chi_{\kappa_\nu}^{\mu_\nu}
\end{pmatrix}
\right\}.
\end{align}
The products of the two-component spinors are given as
\begin{align}
  \chi_\kappa^{\mu\dagger}\left(\hat{r}\right)\chi_{\kappa'}^{\mu'}\left(\hat{r}\right)=&\frac{1}{\sqrt{4\pi}}\sum_{L,M}\left(j_\kappa,-\mu,j_{\kappa'},\mu'|L,M\right)(-1)^{1/2-\mu}S_\kappa
  S_{0LL}(\kappa,\kappa')Y_{LM}\left(\hat{r}\right), \\
  \chi_\kappa^{\mu\dagger}\left(\hat{r}\right)\sigma^i\chi_{\kappa'}^{\mu'}\left(\hat{r}\right)=&\frac{-1}{\sqrt{4\pi}}\sum_{J,L,M}\left(j_\kappa,-\mu,j_{\kappa'},\mu'|J,M\right)(-1)^{1/2-\mu}
  S_\kappa S_{1LJ}(\kappa,\kappa')\left[Y_L\left(\hat{r}\right)\otimes\left(\epsilon^i\right)\right]_{JM},
\end{align}
where $\epsilon$ is a unit vector.
Using these relations, we obtain the effective Hamiltonian Eqs.~\eqref{eq:Heff_beta} and \eqref{eq:xi}.
According to the Wigner--Eckart theorem, the $M$-dependence of the spherical tensor $\Xi_{JLM}$, Eq.~\eqref{eq:xi}, is known as
\begin{align}
  \Braket{f|\Xi_{JLM}\left(\kappa_e,\kappa_\nu\right)|i}=
  \frac{\left(J_i,s_i,J,M|J_f,s_f\right)}{\sqrt{2J_f+1}}
  \Braket{f||\Xi_{JL}\left(\kappa_e,\kappa_\nu\right)||i},
\label{eq:Wigner-Eckart}
\end{align}
where the reduced matrix element $\Braket{f||\Xi_{JL}\left(\kappa_e,\kappa_\nu\right)||i}$ is independent of $M$, and
$J_f$ is the angular momentum of the final nuclear state.

With the obtained Hamiltonian $H_{\mathrm{eff}}$, the decay rate is given by
\begin{align}
\Gamma&=\frac{1}{2J_i+1}\sum_{s_i}\sum_{s_f,s_e,s_\nu}\int\frac{d^3p_\nu}{\left(2\pi\right)^3}\frac{d^3p_e}{\left(2\pi\right)^3}(2\pi)\delta\left(E_\nu+E_e-E_0\right)\left|H_{\mathrm{eff}}\right|^2 \nonumber\\
&=\frac{G_F^2V_{ud}^2\left(4\pi\right)^3}{4\left(2\pi\right)^5}\int_{m_e}^{E_0}dE_ep_eE_e\left(E_0-E_e\right)^2\frac{1}{\left(2J_i+1\right)\left(2J_f+1\right)}\sum_{s_f,s_i}\sum_{s_e,s_\nu}\int d\Omega_ed\Omega_\nu \nonumber\\
&\times\left|\sum_{el}\sum_{neu}\sum_{J,L,M}(-1)^{1/2-\mu_e}\left(j_{\kappa_e},-{\mu_e},j_{\kappa_\nu},\mu_\nu|J,M\right)\left(J_i,s_i,J,M|J_f,s_f\right)\Braket{f||\Xi_{JL}\left(\kappa_e,\kappa_\nu\right)||i}\right|^2,
\label{eq:gamma}
\end{align}
where $J_i$ is the angular momentum
of the initial nuclear state.
For an arbitrary function $X\left(\kappa,\mu\right)$, we have
\begin{align}
\int d\Omega_e\sum_{s_e}\left|\sum_{el}X\left(\kappa_e,\mu_e\right)\right|^2=\sum_{\kappa_e,\mu_e}\left|X\left(\kappa_e,\mu_e\right)\right|^2,
\label{eq:sum_of_sum_el}
\end{align}
and
\begin{align}
\int d\Omega_\nu\sum_{s_\nu}\left|\sum_{neu}X\left(\kappa_\nu,\mu_\nu\right)\right|^2=2\sum_{\kappa_\nu,\mu_\nu}\left|X\left(\kappa_\nu,\mu_\nu\right)\right|^2.
\label{eq:sum_of_sum_neu}
\end{align}
Applying Eqs.~\eqref{eq:sum_of_sum_el} and \eqref{eq:sum_of_sum_neu}, and the orthonormal relation of the Clebsh--Gordan coefficients to Eq.~\eqref{eq:gamma}, one can perform all the angular integral and summation. We then arrive at Eq.~\eqref{eq:decay_rate}.

\section{Explicit formula for the uniform charge distribution \label{app:uniform}}

For the uniform charge distribution of nuclei with radius $R_A$, 
the Coulomb potential for an electron is given as 
\begin{align}
  V_C(r) & = - \frac{\alpha Z}{2R_A}\left[\theta(1-x)(3 - x^2)
    + \theta(x-1)\frac{2}{x}\right],
\end{align}
where $\alpha$ is the fine structure constant and $x=r/R_A$. 
The electron wave functions $D^{(i)},d^{(i)},H^{(2)}$, and $h^{(2)}$ are given as
\begin{align}
  \frac{r}{R}D_k^{(1)}(r) & =  r\left[ \frac{E_e}{2k+1} +\xi s_1(x)\right], \\
  \frac{r}{R}d_k^{(1)}(r) & =  r\left[\frac{m_e}{2k+1}\right], \\
  H^{(2)}(r) & =  r^2\left[- \frac{p_e^2}{2(2k+1)} + E_e \xi s_2(x) + \xi^2 t_2(x)\right],
    \\
  h^{(2)}(r) & =  r^2[  m_e \xi h_2(x)], \\
  \frac{r}{R}D_k^{(3)}(r) & =  r^3\left[- \frac{p_e^2 E_e }{2(2k+1)(2k+3)} +
  p_e^2 \xi s_3(x) + m_e^2\xi t_3(x) + E_e\xi^2 w_3(x) + \xi^3 y_3(x)\right],\\
\frac{r}{R}d_k^{(3)}(r) & =  r^3\left[- \frac{p_e^2 m_e}{2(2k+1)(2k+3)} +
  m_eE_e\xi t_3(x) + m_e\xi^2 z_3(x)\right], 
\end{align}
where $\xi  =  \alpha Z/(2R_A)$ for an electron
and $\xi  =  -\alpha Z/(2R_A)$ for a positron.

The functions $s_a, t_a, w_a, y_a,$ and $z_a$ for $k=1$ are given as
\begin{align}
    s_1(x) &= \theta(1-x) \left(1 - \frac{x^2}{5}\right) + \theta(x-1)\frac{1}{x}\left(1 - \frac{1}{5x^2}\right),\\
    s_2(x) &= \theta(1-x) \left(- 1 + \frac{2 x^2}{15}\right)
    + \theta(x-1)\frac{1}{x}\left(- \frac{5}{3} + \frac{1}{x} - \frac{1}{5x^2}\right),\\
    t_2(x) &= \theta(1-x) \left(-\frac{3}{2}+\frac{2x^2}{5}-\frac{x^4}{30}\right) +
    \theta(x-1)\frac{1}{x^2}\left(- \frac{14}{15} - \frac{1}{5x^2} - 2\ln x\right),\\
    w_2(x) &= \theta(1-x) \frac{x^2}{30} +
    \theta(x-1)\frac{1}{x}\left( \frac{1}{3} - \frac{1}{2x} + \frac{1}{5x^2}\right),\\
    s_3(x) &= \theta(1-x) \left(- \frac{3}{10}+ \frac{3 x^2}{70}\right) +
    \theta(x-1)\frac{1}{x}\left( -\frac{1}{2} + \frac{1}{3x}
                             - \frac{1}{10x^2} + \frac{1}{105 x^4}\right),\\
    t_3(x) &= \theta(1-x) \left(- \frac{1}{5}+ \frac{x^2}{42}\right) +
    \theta(x-1)\frac{1}{x}\left( -\frac{1}{3} + \frac{1}{6 x}
    -   \frac{1}{105 x^4}\right),
     \\
    w_3(x) &= \theta(1-x) \left(- \frac{9}{10}+ \frac{9 x^2}{35} - \frac{x^4}{54}\right)\notag\\ &+
    \theta(x-1)\frac{1}{x^2}\left( -\frac{6}{5} + \frac{1}{x}
    -  \frac{3}{5x^2} + \frac{131}{945 x^3}
       - \frac{2 \ln x}{3}\right),\\
    y_3(x) &= \theta(1-x) \left(- \frac{9}{10}+ \frac{27 x^2}{70}
       - \frac{x^4}{18} + \frac{x^6}{330}\right)\notag\\ &+
    \theta(x-1)\frac{1}{x^3}\left( \frac{1}{15} - \frac{439}{693 x^2}
    - 2\ln x - \frac{2 \ln x}{5 x^2}\right),\\
    z_3(x) &= \theta(1-x) \left(- \frac{3}{10}+ \frac{x^2}{14}-\frac{x^4}{135}\right) \notag\\ &+
    \theta(x-1)\frac{1}{x^2}\left( \frac{2}{15} - \frac{1}{2 x}
    + \frac{1}{5 x^2}
    - \frac{131}{1890 x^3} - \frac{2 \ln x}{3}\right).
\end{align}    
Similar formulas for $k=2$ are given as 
\begin{align}
  s_1(x) &= \theta(1-x) \left(\frac{3}{5} - \frac{x^2}{7}\right)
     + \theta(x-1)\frac{1}{x}\left(\frac{1}{2} - \frac{3}{70x^4}\right),\\
   s_2(x) &= \theta(1-x) \left(- \frac{3}{5} + \frac{3x^2}{35}\right)
    + \theta(x-1)\frac{1}{x}\left(- \frac{9}{10} + \frac{2}{5x} - \frac{1}{70x^4}\right),\\
    t_2(x) &= \theta(1-x) \left(-\frac{9}{10}+\frac{9x^2}{35}-\frac{x^4}{42}\right) +
    \theta(x-1)\frac{1}{x^2}\left(- \frac{271}{420} - \frac{3}{140 x^4} - \ln x\right),\\
    w_2(x) &= \theta(1-x) \frac{x^2}{70} +
    \theta(x-1)\frac{1}{x}\left( \frac{1}{10} - \frac{1}{10x} + \frac{1}{70x^4}\right),\\
    s_3(x) &= \theta(1-x) \left(- \frac{9}{70}+ \frac{13 x^2}{630}\right) +
    \theta(x-1)\frac{1}{x}\left( -\frac{11}{60} + \frac{2}{25 x}
                             - \frac{1}{140 x^4} + \frac{4}{1575 x^6}\right),\\
    t_3(x) &=  \theta(1-x) \left(- \frac{3}{35}+ \frac{x^2}{90}\right) +
    \theta(x-1)\frac{1}{x}\left( -\frac{2}{15} + \frac{3}{50 x}
                              - \frac{2}{1575 x^6}\right),\\
  w_3(x) &= \theta(1-x) \left(- \frac{27}{70}+ \frac{13 x^2}{105}
                              - \frac{23 x^4}{2310}\right)\notag\\ & +
    \theta(x-1)\frac{1}{x^2}\left( -\frac{943}{2100} + \frac{1}{5 x}
    - \frac{1}{20x^4} + \frac{157}{5775 x^5}
    - \frac{\ln x}{5}\right),\\
    y_3(x) &= \theta(1-x) \left(- \frac{27}{70}+ \frac{13 x^2}{70}
    - \frac{23 x^4}{770} + \frac{x^6}{546}\right) \notag\\ &+
    \theta(x-1)\frac{1}{x^3}\left( -\frac{83}{420} - \frac{87}{2860 x^4}
    - \frac{\ln x}{2} - \frac{3 \ln x}{70 x^4}\right),\\
    z_3(x) &= \theta(1-x) \left(- \frac{9}{70}+ \frac{x^2}{30} - \frac{4 x^4}{1155}\right) \notag\\ &+
    \theta(x-1)\frac{1}{x^2}\left( -\frac{103}{2100} - \frac{1}{20x}
    + \frac{1}{140 x^4} - \frac{157}{23100 x^5} - \frac{\ln x}{5}\right).
\end{align}

\section{Table of the coefficients in $\Xi_{JLM}$}
\label{app:Xi}

For the axial vector space component, the effective operator is expressed as
Eq.~(\ref{eq:coef}):
\begin{align}
  \Xi_{JLM}(\kappa_e,\kappa_\nu)
  = \int d\bm{r} [Y_L(\hat{r})\otimes \bm{A}(\bm{r})]_{JM}[
    c_g G_{\kappa_e}(r)g_{\kappa_\nu}(r)  + c_f F_{\kappa_e}(r)f_{\kappa_\nu}(r) ].
  \nonumber
\end{align}
Tables~\ref{GTcoef.tab} and \ref{SDcoef.tab} list the explicit numbers of the coefficients
  of each $(\kappa_e,\kappa_\nu)$
for the Gamow--Teller and spin-dipole transitions, respectively.

\begin{table}[ht]
  \caption{Coefficients in Eq.~(\ref{eq:coef}) for the Gamow--Teller  transition ($L=0, \Delta J^\pi=1^+$).}
  \centering
\begin{tabular}{ccccc}
\hline\hline
  $(\kappa_e,\kappa_\nu)$ &   $(c_g,c_f)$ &&  $(\kappa_e,\kappa_\nu)$ &   $(c_g,c_f)$ \\ \hline
 $( -1, -1)$ & $( \sqrt{2},   \sqrt{2}/3)$& &  $(-2, 1)$ & $( 4/3,   0)$\\
 $( 1, 1)$ & $( -\sqrt{2}/3,   -\sqrt{2})$& & $( 2,-1)$ & $( 0,   -4/3)$ \\
 $(-2,-2)$ & $(-2\sqrt{5}/3, -2/\sqrt{5})$& &  $( 1,-2)$ & $( 4/3,   0)$ \\
 $( 2, 2)$ & $( 2/\sqrt{5},   2\sqrt{5}/3)$& &  $(-1, 2)$ & $( 0, -4/3)$\\
\hline\hline
\end{tabular}
\label{GTcoef.tab}
\end{table}

\begin{table}[ht]
  \centering
  \caption{Same as Tab.~\ref{GTcoef.tab} but  for the
    spin-dipole transition ($L=1, \Delta J^\pi=0^-, 1^-, 2^-$).}
\begin{tabular}{ccccc}
  \hline\hline
  &&  $J=0$ & $J=1$ & $J=2$  \\
  \hline
  $(\kappa_e,\kappa_\nu)$ &&  $(c_g,c_f)$  &  $(c_g,c_f)$  &  $(c_g,c_f)$  \\ \hline
  $(-1,1)$   &&  $(-\sqrt{2} ,\sqrt{2} )$     & $(2/\sqrt{3} ,2/\sqrt{3} )$     &  \\
  $( 1,-1)$  &&  $(-\sqrt{2} ,\sqrt{2} )$     & $(-2/\sqrt{3} ,-2/\sqrt{3} )$     &  \\ 
  $(-2,2)$   &&  $(2 ,-2 )$ & $(-4\sqrt{2/15} , -4\sqrt{2/15})$     & $(\sqrt{2}/5 ,-\sqrt{2}/5 )$ \\
  $( 2,-2)$  &&  $(2 ,-2)$  & $(4\sqrt{2/15} , 4\sqrt{2/15} )$     & $(\sqrt{2}/5 , -\sqrt{2}/5)$ \\
  $(-2,-1)$  &&             & $(\sqrt{2/3} ,\sqrt{2/3} )$     & $(\sqrt{2} ,\sqrt{2}/5 )$ \\
  $( 2, 1)$  &&             & $(-\sqrt{2/3} ,-\sqrt{2/3} )$     & $(-\sqrt{2}/5 ,-\sqrt{2} )$ \\
  $( -1,-2)$ &&             & $(-\sqrt{2/3} ,-\sqrt{2/3} )$     & $(\sqrt{2} ,\sqrt{2}/5 )$ \\
  $( 1, 2)$  &&             & $(\sqrt{2/3} ,\sqrt{2/3} )$     & $(-\sqrt{2}/5 , -\sqrt{2})$ \\
  \hline\hline
\end{tabular}
\label{SDcoef.tab}
\end{table}

\section{Tables of electron and positron wave functions}
\label{app:wftable}

We provide numerical tables of the four constants
  $\alpha_{1}, \alpha_{-1}, \alpha_{2},$ and $\alpha_{-2}$ needed to construct the electron and
  positron wave functions in this paper. Since these constants
  are strongly dependent on the electron momentum $p_e$ and charge number $Z$ of a nucleus,
we rewrite them to
$L_0, \lambda_2,\mu_1$, and $\mu_2$
according to Ref.~\cite{Schopper:1969jkp}:
\begin{align}
  F(Z,E_e) & =  \alpha_{-1}^2 + \alpha_1^2 = F_0 L_0,\\
  \lambda_2 & =  \frac{\alpha_{-2}^2 + \alpha_2^2}{\alpha_{-1}^2 + \alpha_1^2},\\
  \mu_k & =  \frac{k E_e}{\gamma_k m_e}
  \frac{\alpha_{-k}^2 - \alpha_k^2}{\alpha_{-k}^2 + \alpha_k^2},
\end{align}
where
\begin{align}
  F_0(Z,E) & = 4(2p_e R_A)^{-2(1-\gamma_1)}e^{\pi\nu}
  \left|\frac{\Gamma(\gamma_1+i\nu)}{\Gamma(2\gamma_1+1)}\right|^2,\\
  \gamma_k & =  \sqrt{k^2 - (\alpha Z)^2}, \\
  \nu & =  \frac{\alpha Z E_e}{p_e}
\end{align}
for the uniform nuclear charge distribution with radius $R_A$.
Actually, these variables have a milder momentum and charge dependence
than $\alpha_{\pm 1,2}$. First, we calculate $\alpha_{\pm 1,2}$
by numerically solving the Dirac equation and convert them into 
$L_0, \lambda_2,\mu_1$, and $\mu_2$ at various $p_e/m_e$ and $Z$.
These generated tables are respectively interpolated by assuming the following polynomial function
at three regions, $p_e/m_e=$0.01--1, 1--10, and 10--100:  
\begin{align}
P(p_e/m_e,Z)=\sum_{t=-m}^{n}\left[\sum_{s=-m}^nW_{st}\left(\frac{p_e}{m_e}\right)^s\right]Z^t.
\end{align}
The weights of the polynomial $W_{st}$ are determined by the least-square method.
Finally, we reconstruct $\alpha_{\pm 1,2}$ from $L_0, \lambda_2,\mu_1$, and $\mu_2$.
Since all $\alpha_{\pm 1,2}$ values are positive, the reconstruction can be made easily.
We confirm that the resulting numerical tables
are accurate more than 3--4 digits with $(m,n)=(3,4)$.

For the convenience of a user,
we provide a FORTRAN program code to generate $\alpha_{\pm 1,2}$
with a given $p_e/m_e(=$0.01--100) and $Z(=1$--90)
for $\beta^\mp$ decay as supplemental material.
The nuclear charge radius $R_A$ is set to be $1.2A^{1/3}$ fm with
the nuclear mass number $A$.
To cover stable and neutron-rich unstable nuclei for the $\beta^-$ decay,
a variation of the charge radius can be considered among five options:
$A=2Z, 2.5Z, 3Z, 3.5Z$, and $4Z$.

\bibliographystyle{ptephy}
\bibliography{beta_decay_ref}

\end{document}